# Towards Cost-Effective Storage Provisioning for DBMSs

Ning Zhang [#1], Junichi Tatemura [*2], Jignesh M. Patel [#3], Hakan Hacıgümüş [*4]

[#]Computer Sciences Department, University of Wisconsin-Madison, USA
[1]nzhang@cs.wisc.edu  [3]jignesh@cs.wisc.edu

[*]NEC Laboratories America, USA
[2]tatemura@sv.nec-labs.com  [4]hakan@sv.nec-labs.com

## ABSTRACT

Data center operators face a bewildering set of choices when considering how to provision resources on machines with complex I/O subsystems. Modern I/O subsystems often have a rich mix of fast, high performing, but expensive SSDs sitting alongside with cheaper but relatively slower (for random accesses) traditional hard disk drives. The data center operators need to determine how to provision the I/O resources for specific workloads so as to abide by existing Service Level Agreements (SLAs), while minimizing the total operating cost (TOC) of running the workload, where the TOC includes the amortized hardware costs and the run time energy costs. The focus of this paper is on introducing this new problem of TOC-based storage allocation, cast in a framework that is compatible with traditional DBMS query optimization and query processing architecture. We also present a heuristic-based solution to this problem, called DOT. We have implemented DOT in PostgreSQL, and experiments using TPC-H and TPC-C demonstrate significant TOC reduction by DOT in various settings.

## 1. INTRODUCTION

The move towards cloud computing for data intensive computing presents unique opportunities and challenges for data center (DC) operators. One key challenge that DC operators now face is how to provision resources in the DC for specific customer workloads. The focus of this paper is on one aspect of this vast problem – namely how to provision resources in the I/O subsystem. We note that I/O subsystems are often the most expensive components of high-end data processing systems. For example, in the current highest performing Oracle TPC-C configuration [2], the cost of the storage subsystem is $23.9 million compared to $5.2 million for the remaining server.

To fully understand the challenge, consider the dilemma of a modern DC operator, again only focusing on the I/O subsystem. I/O subsystems have gotten incredibly complicated over the last few years primarily due to the disruptive introduction of flash solid state drives (SSDs). Thus it is common for DCs to have servers that have a rich I/O subsystem with a mix of traditional hard disk drives (HDDs) typically in some RAID configuration, and some SSDs. To make matters worse, since the price and the performance characteristics of these I/O devices vary widely, it is not uncommon to find server configurations that have a diverse I/O subsystems with various types of storage devices. For example, a server may have a RAID HDD subsystem, and a high-end fast but expensive SSD (e.g. Fusion IO), and a low-end slow but cheaper SSD (e.g. a Crucial or Intel SSD). DC operators have to make the decision to purchase the server boxes right upfront, and later have to deal with provisioning these resources on (ever changing) workloads. In addition, multiple different workloads may share resources on the same physical box and provisioning the workload requires taking into account physical constraints such as storage capacity constraints. One dilemma that the DC operator faces in this setting is what resources to provision for specific workloads given this rich (I/O) ecosystem.

The problem that we define and address in this paper is as follows: The DC has a cluster of servers each with a rich I/O subsystem on which a set of customer workloads must be provisioned. Service Level Agreements (SLAs) between the DC provider and the customers provide a contract in terms of what each customer can expect. Typical SLAs (via Service Level Objectives embedded in the SLAs) describe characteristics such as expected performance [14] and expected data availability (e.g. SQL Azure's SLA [4]). Given the SLAs, the goal of the DC provider is to provision enough resources to meet the SLAs, while minimizing the *total operating cost*, so as to maximize their profit.

Notice that the objective here is to minimize the total operating cost (**TOC**). In this paper, we consider the TOC to include the amortized hardware cost (incurred during the initial purchase and amortized over the expected lifespan of that hardware), and the run-time energy costs incurred in powering that hardware when running the workload. Extending our work to include other components to TOC such as the management costs and amortized cost of the actual DC facility is fairly straight-forward (see the extended version of this paper [3] for more details).

Now consider the impact of heterogeneous I/O hardware on the TOC. Different I/O devices have different initial costs, storage capacities, performance, and run-time energy costs. SSDs generally run cooler than HDDs (the energy savings is often an order of magnitude with SSDs), but cost more (often more than 10X for the same storage). SSDs have far better random I/O performance. However, the sequential I/O performance of SSDs is comparable to HDDs (which are often setup in RAID configuration), or could be lower than the sequential performance of HDDs for the same cost [23]. Within the context of our problem statement, if we measure the (TOC) for each byte of storage for each unit time of usage, then different I/O devices have different costs, as shown in the first row of Table 1.

|  | **HDD** | **HDD Raid 0** | **L-SSD** | **L-SSD Raid 0** | **H-SSD** |
|---|---|---|---|---|---|
| TOC/GB/hour (cents) | $3.47 \times 10^{-4}$ | $8.19 \times 10^{-4}$ | $7.65 \times 10^{-3}$ | $9.51 \times 10^{-3}$ | $1.69 \times 10^{-1}$ |
| Sequential Read (ms/IO) | **0.072** (0.174) | **0.049** (0.096) | **0.036** (0.053) | **0.021** (0.037) | **0.016** (0.013) |
| Random Read (ms/IO) | **13.32** (8.903) | **12.19** (2.712) | **1.759** (1.468) | **1.570** (0.826) | **0.091** (0.024) |
| Sequential Write (ms/row) | **0.012** (0.039) | **0.011** (0.034) | **0.020** (0.341) | **0.013** (0.082) | **0.009** (0.025) |
| Random Write (ms/row) | **10.15** (8.124) | **11.55** (3.770) | **62.01** (37.45) | **21.14** (17.71) | **0.928** (0.986) |

Table 1: The Cost and I/O profiles of different storage classes under 1 and 300 degree of concurrency: (1) The first row lists five different storage types/classes that we use in our experiments. These storage types are discussed in more detail in Section 4.1. L-SSD stands for Low-end SSD, H-SSD stands for High-end SSD. (2) The second row shows the storage cost in terms of cents per GB per hour, calculated using the method described in Section 2.1. (3) The remaining four rows show the performance of the storage types on four typical I/O access patterns. In each cell, the boldfaced number is for a workload with a single DB thread, whereas the number in the parentheses is the I/O performance with 300 concurrent DB threads. See Section 3.5 for details about the concurrency parameter.

Thus, provisioning the I/O storage subsystem to minimize the TOC is an optimization problem that considers the range of available I/O devices, examines the capacity constraints for each device, and the performance characteristics of each workload and the I/O devices, to compute a *data layout* that minimizes the TOC, while meeting the SLAs. In this paper, we propose, implement, and evaluate a technique to address this problem.

The contributions of this paper are as follows:

1. We formulate the problem of data placement to minimize the TOC for cloud-hosted DBMSs.

2. We present a practical solution, called DOT, for the data placement problem that can be incorporated in existing DBMSs. The DOT method extends the DBMS query optimizer's cost estimation component to consider the I/O speeds of various storage devices. DOT then exploits the ability of most modern DBMSs to output query plans (without actually executing the plan) that are then fed its TOC optimizing component. DOT's TOC optimizing module uses a novel heuristic to compute a desirable data placement.

3. Finally, we have implemented the DOT method in PostgreSQL, and using TPC-H and TPC-C based workloads to verify its effectiveness, showing that in many cases, the data layout recommended by DOT is up to 5X more cost-effective than other simple layouts.

On a cautionary note, we acknowledge that in this initial paper, we only focus on a small part of the problem of minimizing TOC in DCs. For example, we focus only the I/O subsystem, we have focused on relatively simple workloads, ignored multi-tenancy, and we do not consider dynamic workload migration. The area of minimizing TOC in DBMSs is fairly new, and there are many open unsolved problems – we hope that this work seeds other work in this area to examine and solve these many open problems.

The remainder of this paper is organized as follows: Section 2 introduces our *cost model* and the problem definition. The DOT method is described in Section 3. Section 4 contains experimental results, and Section 5 describes various extensions to our work. Related work is presented in Section 6, and Section 7 contains our concluding remarks.

## 2. PROBLEM DEFINITION

To illustrate the problem of TOC-based storage provisioning, consider the following motivating scenario: *Given a data center with many database workloads, a data center administrator needs to build a database server configuration that consists of various storage devices. A critical question is how to choose the storage devices and how to place data on these devices for each workload. Although it is said that a high-end SSD performs much better than a hard disk drive (HDD), the administrator is not sure if it pays off in terms of the (TOC) cost. The administrator wants to achieve better cost-performance while the performance (e.g., response time) meets the given requirements as set by individual SLAs.*

### 2.1 Cost Model

We note that coming up with a cost (price) model of a storage device is a complex problem as it depends on various factors, such as vendor agreements and volume discounts. In this work, we focus on the following simple model:

**Storage price** *(cent/GB/hour)*: For each storage, the amortized storage cost is calculated and amortized by space and time (cent/GB/hour). Table 1 show our calculated storage prices for five actual devices: (1) HDD, (2) HDD RAID 0, (3) L-SSD, (4) L-SSD RAID 0 and (5) H-SSD. In this calculation, the purchase cost of the I/O device is distributed over 36 months, and the energy cost is computed using a cost of 0.07 per kWh of energy consumption [16].

We model the available storage classes as $D = \{d_1, .., d_M\}$, where each $d_i$ is a specific storage class (e.g. HDDs in RAID 0 configuration). The price of $d_j$ is denoted by $p_j$, and the price vector $P = \{p_1, .., p_M\}$.

**Layout cost** *(cent/hour)*: Assume that a database is laid out on $D$, taking $S_j$ GB space for each class $d_j$ ($S_j \geq 0$). Now, let $L$ denote this particular layout. (We describe how to compute the layout $L$ in Section 3.4). Then, the cost per hour for this layout $L$, denoted as $C(L)$, is computed as $C(L) = \Sigma_{d_j \in D}(p_j * S_j)$.

**Workload cost** (TOC) *(cent/task)*: Assume that the database with layout $L$ achieves a throughput $T(L, W)$ (measured in tasks/hour) for a given workload $W$. Then, the workload cost is defined as $C(L, W) = C(L)/T(L, W)$ (more details are below in Section 2.3). In this paper, we refer to this workload cost as TOC.

Our problem is to find a layout $L$ over $D$ that minimizes $C(L, W)$ for a given workload $W$ under the price model $P$

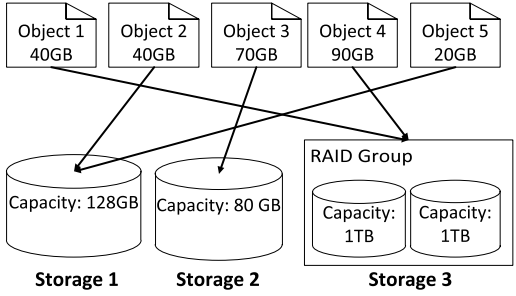

Figure 1: Allocating objects to storage classes

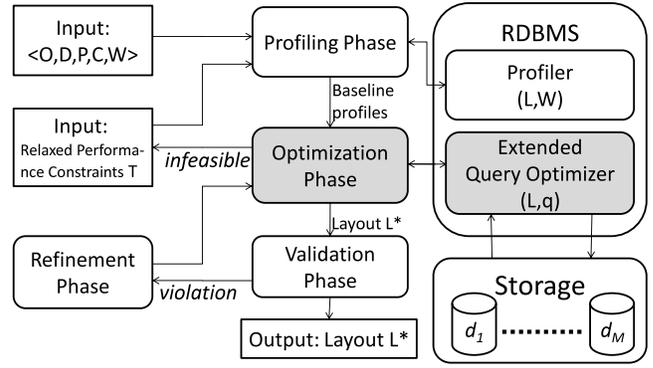

Figure 2: Overview of the DOT method

with constraints on storage capacity and workload performance, as described in the remainder of this section.

## 2.2 Data Layout and Capacity Constraints

We assume that the storage system provides $M$ different classes of storage $D = \{d_1, \cdots, d_M\}$. A storage could be an individual device, or a RAID group, and we use $c_j$ to denote the capacity of the storage class $d_j$.

A database instance consists of a set of objects $O = \{o_1, \cdots, o_N\}$, such as individual tables, indices, temporary spaces or logs, that must be placed on one of the storage classes in $D$. We use $s_i$ to denote the size of the data object $o_i$. (In this paper, we do not consider partitioning or replication of objects, which are important considerations and promising directions for future research.)

A **data layout** $L$ is defined as a mapping from $O$ to $D$, where $L(o)$ indicates the storage mapping for object $o$. Let $O_j$ denote a set of objects laid out on $d_j$, i.e., $O_j = \{o | L(o) = d_j, o \in O\}$. A valid data layout must conform to the **capacity constraint** of each storage, i.e., $\Sigma_{o_i \in O_j} s_i < c_j$ ($j = 1, \cdots, M$). Figure 1 illustrates a sample layout.

## 2.3 Workloads

We model a workload, $W$, as a set of query sequences, $\{[q_1^1, \cdots, q_n^1], \cdots, [q_1^c, \cdots, q_n^c]\}$, where each $q_i^j$ is a database query, and $c$ denotes the concurrency of the workload $W$. Let $t(L, W)$ be the execution time of $W$ under layout $L$. Then, the workload cost (TOC) is $C(L, W) = C(L) * t(L, W)$.

## 2.4 Performance Constraints

In our model, we assume that there are performance related SLA constraints associated with the queries (so there is some limit on the query performance degradation that can be tolerated). These *performance constraints* $T$ can be modeled as the upper bound of each query execution time, $T = \{t_i^j\}$, where $t_i^j$ is the response time cap for query $q_i^j$.

While the framework above uses query response time as the performance metric, this framework can be adapted to consider other performance metrics, such as throughput rate. In fact, in Section 4, we use response time constraints for individual queries for the TPC-H DSS workload, and use throughput constraints for the TPC-C OLTP workload.

In this paper, rather than using an absolute performance constraint, we define the performance constraint as a ratio that is relative to the best performance (similar to the way the measure degradation limit that was used in [26]). For instance, the performance constraint 1/3 means that the workload can be up to 3 times slower than the best case (e.g., when all the objects are placed on a high-end SSD, when a high-end SSD is one of the available storage classes). Using this method of defining the performance constraint, we can demonstrate various cases of cost-performance trade-offs, and compare them to the "best" performing case.

We also note that our framework can be generalized to allow a broader definition of performance constraints, including capturing a general distribution of the performance metric (e.g. must be faster than $x$ seconds in 95% of the cases, and faster than $1.5x$ seconds in 99% of the cases). Such extensions are part of future work.

## 2.5 Problem Formulation

Our problem can be formally stated as follows:

**Input**: (1) Database objects $O = \{o_1, \cdots, o_N\}$, (2) Storage classes $D = \{d_1, \cdots, d_M\}$ with price (TOC/GB/hour) $P = \{p_1, \cdots, p_M\}$ and capacity $C = \{c_1, \cdots, c_M\}$, (3) Query workload $W = \{[q_1^1, \cdots, q_n^1], \cdots, [q_1^c, \cdots, q_n^c]\}$ with performance constraints $T = \{t_i^j\}$.

**Output**: A layout $L : O \to D$ that minimizes the TOC $C(L, W) = C(L) * t(L, W)$ for a given $W$ where

$$C(L) = \Sigma_{d_j \in D}(p_j * S_j)$$

under the capacity constraints, $\Sigma_{o_i \in O_j} s_i < c_j$ ($j = 1, \cdots, M$), and performance constraints $T = \{t_i^j\}$.

## 3. THE DOT METHOD

A naive way to solve the layout optimization problem (formulated in Section 2.5) is to enumerate and validate each possible layout candidate. However, this approach is computationally expensive, since with $M$ different storage classes and $N$ data objects, the number of all possible data layouts is exponentially large, i.e. $M^N$.

Our method to compute a **D**ata layout **O**ptimized to reduce the **T**OC, called **DOT**, is shown in Figure 2. There are four steps/phases in our solution: profiling, optimization, validation and refinement.

The technique starts by profiling the workloads on some baseline layouts, $L_1, \cdots, L_k$, to generate a number of *workload profiles* that are then used in an optimization phase. Briefly, a workload profile models the I/O behavior of the workload when it runs on a baseline layout (e.g. for the query `select count(*) from` $A_i$ `where id > A and id < B`, it estimates how many random and sequential read I/Os are incurred on the table $A_i$ when the table $A_i$ and its indices are placed using some specific layout.) We discuss the profiling phase and the baseline layouts in detail in Section 3.4.

Then, in the optimization phase, we employ an heuristic approach that makes use of the workload profiles and the workload performance estimates from an *extended DBMS*

**Procedure 1** Optimization Phase of DOT

**Input:** DOT input $<O, D, P, C, W, T>$, workload profile $X$
**Output:** Layout $L^*$

    $L \leftarrow L_0$, $L^* \leftarrow L$
    $c^* \leftarrow estimateTOC(W, L)$
    $\Delta \leftarrow enumerateMoves(O, D, P, X)$
    **for** $i = 0 \rightarrow |\Delta|$ **do**
        $m \leftarrow \Delta[i]$, $L_{new} \leftarrow m(L)$
        $(c, T') \leftarrow estimateTOC(W, L_{new})$
        **if** $feasible(\{L_{new}, C\}, \{T', T\})$ **then**
            $L \leftarrow L_{new}$
            **if** $c < c^*$ **then**
                $L^* \leftarrow L$, $c^* \leftarrow c$
            **end if**
        **end if**
    **end for**

*query optimizer* to explore the space of possible data layouts. This optimization phase outputs a recommended layout $L^*$ that satisfies all the constraints (See Section 2.5). We describe this heuristic optimization approach in Section 3.1. The extended DBMS query optimizer has a new cost estimation module that considers the different I/O speeds of storage devices to give more precise estimates. We discuss how to extend the query optimizer in Section 3.5.

The (heuristic) method used in the optimization phase is not guaranteed to output a feasible layout; and rather than returning a recommended layout, it may return an answer marked as "*infeasible*," which may mean that the process missed a feasible layout that exists (i.e., false negative), or that there is no feasible layout since the performance constraints are too strict. In either case, the performance constraints must be relaxed in order to compute a layout. The third phase, namely the validation phase, checks if the recommended layout really confirms to the performance constraints through a test run of the workloads on the recommended layout. If the test run "fails", then the system goes to the refinement phase. This refinement phase uses real runtime statistics (such as the actual numbers of I/O incurred in the test run, buffer usage statistics, etc.), and uses those as the input (instead of going to the profiling phase) to redo the optimization phase. In the interest of space, we do not discuss the refinement phase in detail in this paper.

### 3.1 The Heuristic Approach used in DOT

The pseudocode for the heuristic optimization module in DOT is shown in Procedure 1. This procedure enumerates the layout candidates and returns the layout, $L^*$, that has the minimum estimated TOC (i.e., $C(W, L)$) amongst all the candidates. The challenge here is how to enumerate a promising subset of the possible layouts.

Our basic approach is to (1) start from a layout $L_0$ that places all the objects on the most expensive storage class (say, $d_1$), and (2) gradually move objects from $d_1$ to other less expensive storage classes as long as the new layout $L_{new}$ and its estimated performance $T'$ satisfies the capacity constraints $C$ and the SLA constraints $T$ (checked by procedure $feasible$ in the pseudocode). Notice that, in our approach, the move candidates, $\Delta$, are generated *only once* at the beginning of the procedure and are applied one by one, yielding $|\Delta|$ layouts to be investigated.

The key component of this procedure is to generate $\Delta$, a sequence of object moves. For each iteration, a move $m$ in $\Delta$ is applied to a layout $L$, resulting in a new layout $m(L)$. Here, as a heuristic, we want to apply a more beneficial move (i.e., larger TOC reduction) earlier. The subprocedure *enumerateMoves* should generate move candidates in such a promising order (we provide the pseudocode later in Procedure 2), which we achieve by using a heuristic function (Section 3.3) to assign a priority score for each move. This function considers the impact of a move that comprises of a layout cost reduction and a workload performance penalty. The performance penalty is estimated based on the estimated I/O time over the objects. After sorting the move candidates in terms of their priority scores, we apply them in sequence to generate new candidate layouts.

A simple method to generate a set of move candidates is to move an object $o \in O$ to a storage class $s \in D$ one by one, as was done in [10]. In this case, the sub-procedure *enumerateMoves* would generate $M$ moves for each object. By applying the moves one by one, DOT would investigate $O(MN)$ layouts. However, this approach has a serious limitation as it ignores the interactions between the objects. Since the move of one object can significantly change the I/O access pattern of another object, by ignoring the interaction between the objects, this simple approach ignores the change in performance (e.g. the amount of I/O time) and indirectly affects the calculations of priority scores.

A notable example of such interaction between objects is the interaction between a table and its index: Assume that a table has an index (e.g. B+ tree) on its primary key, and a query wants to retrieve records in a given range on its primary key (e.g., `select * from table A where A.id > 10 and A.id < 1000`). Now, consider placing the index on either an SSD or a HDD, and the performance of these two placement strategies. In this case, the performance of the two placement methods will depend on the the placement of the base table. For instance, when the table is on the HDD, the query planner may choose to only use a *sequential scan* on the table to execute the query. In this case, the placement of the index has *no* impact to the I/O cost since it is not accessed *at all*. However, if the table is placed on the SSD, placing the index on the SSD may let the query planner choose an *index scan* to access both the table and the index for greater performance, as this plan leverages the SSD's faster random I/O speed. Thus, we should not ignore the interaction between objects, e.g. a table and its index.

Our heuristic approach is to put objects into groups, referred to as *object groups*, and consider interaction only *within* a group: we put a table and its indices in a group and consider all the combinations of their placements on different storage classes. For example, in the case of a table with one index, and only two devices, a HDD and a SSD, we consider (1) placing both the table and its index on the HDD device; (2) placing the table on the SSD device and the index on the HDD device; (3) placing the table on the HDD device and the index on the SSD device, and (4) placing both the table and its index on SSD device.

On the other hand, in our heuristic approach, we assume independence between objects *across* different groups.

Procedure 2 shows the pseudo code of *enumerateMoves*, which employs the idea of object groups. The high-level description of the procedure is as follows (see Section 3.2 for details): Data objects $O$ are classified into groups $G$. For each group $g$ in $G$, all the placement combinations of objects

277

**Procedure 2** enumerateMoves: Enumeration of moves
**Input:** $<O, D, P, X>$
**Output:** a list of moves $\Delta$

$G \leftarrow grouping(O), \Delta \leftarrow \phi, \Sigma \leftarrow \phi$
**for all** $g \in G$ **do**
  **for all** $p \in D^{|g|}$ **do**
    $m \leftarrow move(g, p)$
    $s \leftarrow score(m, X, D, P)$
    $\Delta \leftarrow append(\Delta, m), \Sigma \leftarrow append(\Sigma, s)$
  **end for**
**end for**
$\Delta \leftarrow sort(\Delta, \Sigma)$

---

in $g$ over storage classes $D$ are considered, and a move $m$ is generated for each combination. $\Delta$ is a list of such moves sorted in the order of priority.

Next, we describe how move candidates are enumerated based on object groups (Section 3.2), the priority score of move candidates (Section 3.3), and workload profiles that are used to calculate the priority score (Section 3.4).

### 3.2 Object Groups

We divide the database objects in $O$ into *object groups* so that interactions between objects in a group is higher for objects within a group than objects in different groups. We assume that any performance gain (or loss) due to a move (from one storage class to another) is independent between objects in different groups. Let us represent a group of objects as a vector $g = (o^1, \cdots, o^K)$. Then, the placement of a group can also be represented as a vector $p = (d^1, \cdots, d^K) \in D^K$. The number of possible placements of a group is $O(M^K)$, where $K$ is the size of the group.

The move of a group $g$ to $p$ is denoted as $m(g, p)$. As shown in Procedure 2, *enumerateMoves* considers all the possible moves $m(g, p)$. The size of $\Delta$ is thus $O(GM^K)$ where $G$ is the number of groups and $K$ is the size of a group ($N = GK$).

While in the current version of DOT, a group consists of the table and its indices, in general, we could introduce other grouping to capture further interactions. However, we need to carefully choose a grouping scheme so that the size $K$ does not become too large. Notice that, if we put all the objects in one group to consider all interactions, our algorithm becomes an exhaustive search method to enumerate all the $O(M^N)$ layouts.

In the current grouping scheme, $G$ is the number of tables and $K$ is as large as the number of indices on each table. Since in many practical cases $K$ is likely to be far smaller than $G$, so the number of layout candidates $O(GM^K)$ in DOT is much smaller than $O(M^{KG})$ (i.e., exhaustive search).

### 3.3 Priority Score

In Procedure 2, a priority score $s$ for a move $m$ is calculated using workload profile $X$ and storage information $(D, P)$. The priority score is derived from two components: *performance penalty* and *layout cost saving*.

First, we describe the notion of a *performance penalty* that estimates the impact of a move $m$ relative to the workload performance. The performance penalty is described using a term called the *I/O time share*, which is the accumulated I/O time over objects $o$ in $g$.

We use the following four types of I/Os to model the typical DBMS query I/O access pattern [10]: sequential read (SR), random read (RR), sequential write (SW) and random write (RW). Now, let $R$ denote the set of these I/O types. As shown in Table 1, we are provided with the time of one I/O operation $\tau_r^d$ for each type $r \in R$ and storage $d \in D$. From this information, we need to estimate the accumulated number of I/O operations on $o$.

We use the profiling phase to estimate the number of I/O operations for each object (Section 3.4). As we have discussed above, the number of I/O operations on a specific object can be very different depending on the placement of not only this object but also other objects in the same group. Thus, we estimate $\chi_r^p[o]$, the number of I/O of type $r$ on $o$ when the group $g$ is placed in a specific placement $p$.

Based on the workload profiles $X = \{\chi_r^p[o]\}$, we estimate the *I/O time share* of an object group $g$ when it is placed in $p$:

$$\mathcal{T}^p[g] = \sum_{o \in g} \sum_{r \in R} \chi_r^p[o] * \tau_r^{p[o]} \quad (1)$$

Here, $p[o]$ is the storage class assigned by the placement $p$ for the object $o$.

Then, the *performance penalty* of a move $m(g, p)$ from the initial layout $L_0$ can be defined as follows:

$$\delta_{time}[m] = \mathcal{T}^p[g] - \mathcal{T}^0[g] \quad (2)$$

Now consider the component the *layout cost saving*, which estimates the impact of a move $m$ on the layout cost $C(L)$. Let $m(L)$ be the layout given by applying $m$ to $L$. Then, the *cost saving* of a move $m$ is:

$$\delta_{cost}[m] = C(L_0) - C(m(L_0)) \quad (3)$$

The definition of $C(L)$ is given in Section 2.1.

Finally, the priority score of a move $m$, denoted as $\sigma[m]$, is defined by considering both the *performance penalty* and the *layout cost saving*, and is calculated as:

$$\sigma[m] = \delta_{time}[m]/\delta_{cost}[m] \quad (4)$$

The procedure *enumerateMoves* sorts all possible moves $m(g, p)$ by their scores in the ascending order.

### 3.4 Workload Profiles

The objective of the profiling phase is to measure the I/O behavior of the workload when an *object group* $g$ is laid out using the placement $p$. This phase produces several workload profiles, where each profile corresponds to a specific placement. As discussed above, the placement $p$ of an object group can impact the optimizer's choice of query plans, resulting in very different I/O costs/profiles. Thus, when we profile the workload, we consider these object interactions by enumerating all possible placements of an object group.

A lightweight method to enumerate all possible placements of the object groups is to use a small set of layouts, referred as *baseline layouts*. For instance, consider a case where each table has only one index on the primary key. Then we have a set of object groups of size 2 (i.e., $K = 2$). For each group, we want to measure I/O profiles for all the $M^2$ placement patterns. To do this, we use the $M^2$ baseline layouts $\{L_{(i,j)} : 1 \leq i, j \leq M\}$ defined as follows: $L_{(i,j)}$ places all the tables on $d_i$ and all the indices on $d_j$. That is, each group object has the same placement $p$, where $p = (d_i, d_j)$. In general, we have $O(M^K)$ baseline layouts where $K$ is the (maximum) size of an object group. Compare to the number of all possible layouts that cover all the combinations amongst different groups (which is again $O(M^N)$), profiling the workloads on the baseline layouts has a lower total complexity when $K \ll N$. Notice that, by using only the baseline layouts in this manner, we assume independence of the placements across different groups, which is the same assumption we made for our heuristics.

A workload profile on a baseline layout, $L_p$, consists of the number of I/O in terms of the I/O types and the data objects. Here, $\chi_r^p[o]$ is given as the number of I/Os of type $r$ on object $o$ when the workload is executed over $L_p$. The workload profiles can be calculated either through (a) an estimate computed by our extended query optimizer as described in Section 3.5, or (b) a sample test run of the workload on $L_p$. (We see both cases in the results described in Sections 4.4 and 4.5 respectively.)

We also notice that there is an opportunity to prune the baseline layouts that are being profiled. If we can infer that the query optimizer will choose the same plans on layouts $L_p$ and $L_q$, we only need to profile one of these. In Section 4.5, we show a special case where only one layout is profiled. A general pruning method, however, is an open issue.

## 3.5 Extended Query Optimizer

The heuristic step in DOT (described above in Section 3.1) estimates the TOC and then checks the performance constraint for a candidate layout by calling the query optimizer's estimation module to estimate the performance of the workload for that layout. To enable this check, the query optimizer should support, or has to be extended to support: (1) query plan optimization that is aware of the I/O profiles of different storage classes; (2) execution time estimation of the derived query plan. In this paper, we have extended the open source RDBMS, PortgreSQL, to accommodate these requirements of the DOT framework.

A typical RDBMS such as PostgreSQL does not consider different I/O performances for heterogeneous storage classes. However, as we have discussed, the best query plan can depend on the specific data layout. For example, the choice between a nested-loop join using an index (indexed NLJ) and hash join (HJ) given specific selectivities depends on the random versus the sequential I/O performance characteristics of the different storage classes. In other words, if we change the data layout, the cheapest query plan may also change, and we need make the optimizer aware of this interaction. To do that, we incorporate I/O profiles (as seen in Table 1) into the query plan cost estimation module.

Next, we introduce a module that estimates the query response time. The PostgreSQL optimizer can output a query plan without actually executing the query. This plan includes statistics, such as the query plan cost, the number of I/Os for a scan and the number of rows processed by query operators (e.g., hashing or sorting). We utilize these statistics to estimate the *I/O time* associated with executing a query, and use the *CPU time* estimates already provided by the query optimizer to approximate the query response time as the sum of these two components. Methods for estimating the *CPU time* in this setting are well known [26], and here we only focus on estimating the *I/O time*.

For simplicity, we do not analyze the effect of cached data in the buffer pool, which can significantly reduce the number of actual I/O in the query. We also ignore the cost of actually outputting the results.

|  | **HDD** | **L-SSD** | **H-SSD** |
|---|---|---|---|
| Brand & model | WD Caviar Black | Imation M-Class 2.5" | Fusion IO ioDrive |
| Flash type | N/A | MLC | SLC |
| Capacity | 500GB | 128GB | 80GB |
| Interface | SATA II | SATA II | PCI-Express |
| RPM | 7200 | N/A | N/A |
| Cache Size | 32MB | 64MB | N/A |
| Purchase cost | $34 | $253 | $3550 |
| Power | 8.3 Watts | 2.5 Watts | 10.5 Watts |

**Table 2: Storage class specifications**

Instead of using the I/O performance numbers of the devices published by the manufacturer or as seen from the OS level, we benchmark the *effective* I/O performance of each I/O request as observed by the DBMS, since: (1) with this approach, various overheads (e.g. latch overhead) and benefits (e.g. DB buffers) are incorporated. (2) we can model the influence of concurrent DB queries on I/O performance.

Here we use the term *degree of concurrency* to refer to the number of concurrent DBMS query processing threads/processes, and use this concept to model how the I/O subsystem behaves when there are concurrent queries.

### 3.5.1 Benchmarking the I/O Characteristics

In general, our method of benchmarking the storage classes follows the profiling method used in [10]. However, we generalize their method for benchmarking the storage class under certain concurrency: we concurrently run $K$ threads that issue queries over their own tables, i.e., thread $i$ issues a query to table $A_i$. Each table has a primary key id which is indexed with a B+ Tree.

**Read I/O**: For read queries, we use a `count(*)` query so as to minimize the costs associated with producing the output. Each thread issues the following queries:

- **Sequential Read (SR)**: Each thread issues the following query: `select count(*) from` $A_i$.

- **Random Read (RR)**: Each thread issues a sequence of queries, using the template: `select count(*) from` $A_i$ `where id = ?`, with randomly selected `id` values.

The time for each I/O is calculated by dividing the total elapsed time of running all queries by the total number of read I/O requests.

**Write I/O**: We benchmark the write I/O characteristics based on the I/O characteristics that is observed end-to-end from inside the DBMS, and is estimated as follows:

- **Sequential Write (SW)**: The SW performance is measured by having each thread issue a large number of insertion queries, where each query inserts a single row using the template: `insert....into` $A_i$.

- **Random Write (RW)**: To measure the RW performance, each thread issues a sequence of update queries using the template `update` $A_i$ `set a = ? where id = ?` with randomly selected `id` values. Notice that an update query consists of random read and random write. To estimate RW from update queries, we subtract the RR I/O time (as estimated above) from the total RW execution time. See [3] for more details.

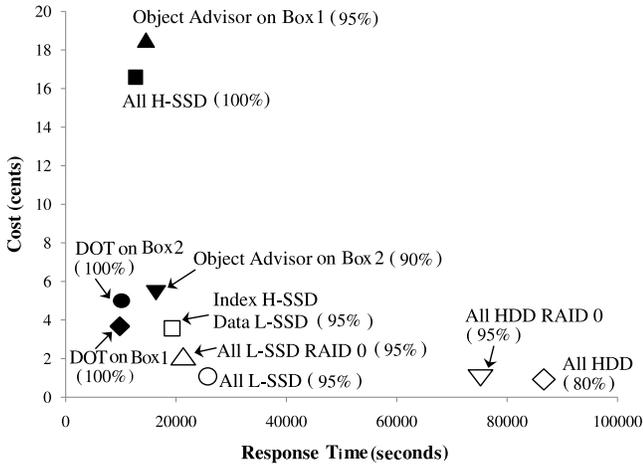

Figure 3: The original TPC-H workload with relative SLA = 0.5. The number in parenthesis associated with each label indicates the PSR value (%).

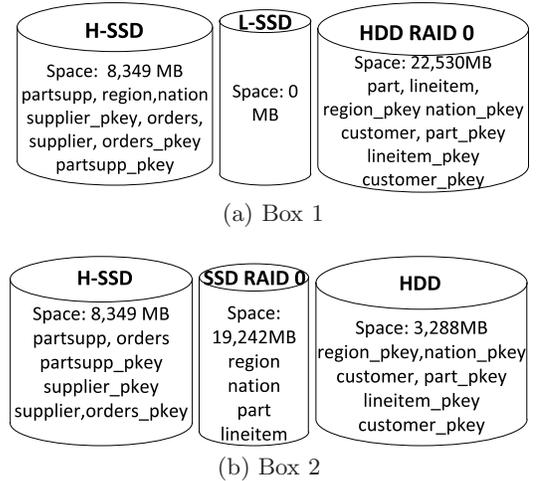

Figure 4: Data layouts with relative SLA = 0.5 and the original TPC-H workload.

Again, the time for each write request (i.e. per row) is calibrated by dividing the total elapsed time by the total number of queries.

Table 1 shows the results from running this benchmark on each storage class that we use in our evaluation (below), with degree of concurrency values of 1 and 300. In our experiments described below, we use values with concurrency 1 for the DSS workloads and 300 for the OLTP workloads.

We note that our DOT framework simply needs some method to characterize the I/O devices. The method described above is simply what we used in our evaluation in this paper, and can be substituted with any other method without impacting the generality of our DOT framework.

## 4. EXPERIMENTAL RESULTS

In this section, we experimentally evaluate our layout technique using an implementation of DOT in PostgreSQL, and demonstrate the effectiveness of our methods using both the TPC-H benchmark (to represent a DSS workload) and the TPC-C benchmark (to represent an OLTP workload).

### 4.1 Hardware and Software Specifications

Our experimental platform is a server system with a 2.26 GHz Intel(R) Xeon CPU E5520 with 8 cores and 64GB ECC memory. To allow experiments in parallel and to avoid having to swap I/O devices for each experiment, we actually used two machines that were identical (same CPU, motherboard, memory, etc.), but had separate storage subsystems. These two storage subsystems are:

- **Box 1:** one HDD RAID 0, one L-SSD and one H-SSD.
- **Box 2:** one HDD, one L-SSD RAID 0, and one H-SSD.

DOT is performed for each box individually, resulting in two separate recommendations. For instance, DOT on Box1 recommends a layout given the 3 storage classes HDD RAID 0, L-SSD, and H-SSD as part of the input.

The specifications of the HDD, the L-SSD and the H-SSD is shown in Table 2. RAID 0 is implemented using **two** identical storage devices and a Dell SAS6/iR RAID controller. This controller costs $110 and has a 256MB onboard cache.

The *storage price* for these storage classes, shown earlier in Table 1, is calculated from the amortized cost (over 36 months) of its purchase cost (including the RAID controller if needed) and the $0.07kWh data center energy cost [16]. The power dissipation in Table 2 is derived from the average values for read and write operations for each storage device. Also, the power surcharge of the RAID controller is 8.25W.

The server runs CentOS (Linux kernel 2.6.18) and PostgreSQL 9.0.1, with our *extended query optimizer* (as discussed in Section 3.5) and the I/O profiling table shown in the Tables 1. We set the PostgreSQL shared buffer to 4GB. In addition to the storage subsystems described above, each machine had an additional 500GB disk that holds the OS, DBMS binaries, and the database log files. Finally, OS caching is turned off for both the log files and the data disks.

### 4.2 Simple Layouts

We use the following "simple" layouts to compare with the layouts that are recommended by DOT:

- **All H-SSD:** All objects placed in the H-SSD (i.e., $L_0$)
- **All L-SSD RAID 0:** All objects placed in the L-SSD RAID 0
- **All L-SSD:** All objects placed in the L-SSD
- **All HDD RAID 0:** All objects placed in the HDD RAID 0
- **All HDD:** All objects placed in the HDD
- **Index H-SSD Data L-SSD:** Indexes in the H-SSD and Data in the L-SSD.

We have also implemented the **Object Advisor (OA)** [10] method in PostgreSQL, as OA is the closest previously known method to DOT. We note that OA optimizes only for workload performance and not the TOC.

### 4.3 Performance Metrics

As discussed in Section 2.4, following the methodology in [26], as a performance measure we use a metric called the "relative SLA", which is the performance for a workload with a given data layout compared to the performance of that workload with all the data on the H-SSD (which is typically the highest performing case). For instance, relative SLA = 0.5 implies that the target performance SLA is half of the performance with all the data on the H-SSD. For the target performance metrics, we use the response time of each query for the TPC-H workload and the total throughput for the TPC-C workload.

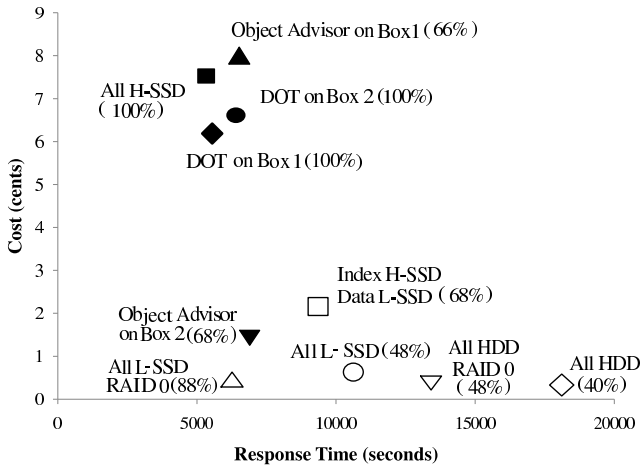

Figure 5: The modified TPC-H workload with relative SLA = 0.5. The number in parenthesis associated with each label indicates the PSR value (%).

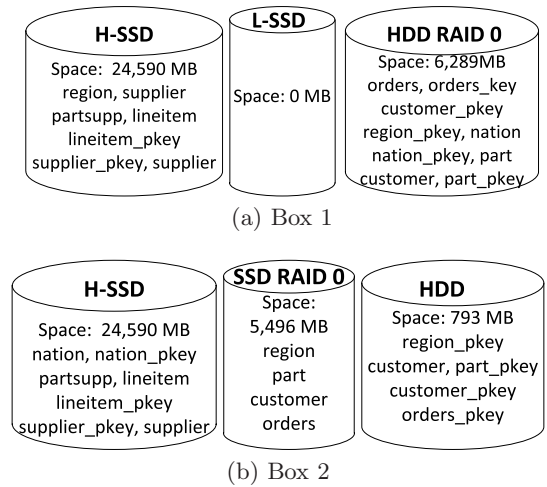

Figure 6: The advised data layouts with relative SLA = 0.5 and the modified TPC-H workload.

Notice that a simple layout, which is not aware of SLA, can fail to meet the target performance. We need an overall measure to indicate the degree of SLA violation of such a layout. For the TPC-H workload, we measure the fraction of the queries that don't meet their relative SLA, using a ratio called the *performance satisfaction ratio* (PSR). For example, a PSR value of 75% means that 75% of queries in the workload meet their relative SLAs and 25% of them do not. For the TPC-C workload, we do not need an additional measure since the throughput performance itself serves as such an indicator.

## 4.4 TPC-H Experiments

For experiments on the DSS workloads, we used two flavors of the TPC-H workloads. These workloads are:
**The original TPC-H workload:** Following the methodology in [22], we use 66 queries generated from the original 22 TPC-H query templates as this workload. Thus in this workload, each TPC-H query occurs three times in the mix. The workload is executed sequentially with the SR I/O as the dominating I/O type.
**A modified TPC-H workload:** We use the exact five TPC-H query templates (Query # 2, 5, 9, 11, 17) that were used in [10]. These five queries are modified in the same manner as in [10] to simulate an Operational Data Store environment. The modifications to these queries is to add more predicates (on the `part` key, `order` key and/or the `supplier` key) to the "where" clause of the queries, so as to reduce the number of rows that are returned. As a result, this workload now has a mix of random and sequential read I/O (Mixed I/O). This workload has a total of 5 query templates that are executed sequentially 20 times, to produce a workload with a 100 queries. The actual queries that we used in this experiment can be found in [3].

In these experiments, we vary the relative SLA (to values of 0.5 and 0.25) without setting any capacity limits on the storage classes.

In addition, we evaluate the heuristics in DOT by comparing it with an exhaustive search approach. For these experiments, we use a smaller workload (for exhaustive search to be tractable) and vary the capacity limits (to make it more challenging for the DOT heuristics).

For all the TPC-H experiments, a 30GB TPC-H database is generated (scale factor 20) and all the tables are randomly reshuffled so that they are not clustered on the primary keys.

### 4.4.1 The Original TPC-H Workload

Figure 3 shows the cost/performance comparison amongst the different layouts when the relative SLA is set to 0.5. The response time is the time to complete the workload and the cost is the measured TOC. The corresponding PSR values are shown in parenthesis in the figure. So, for example, the PSR value for the All L-SSD case is 95%.

From Figure 3, we make the following observations: First, our heuristic layouts on Box 1 and Box 2 produce significant savings — more than **3X** — in terms of the TOC against the All H-SSD layout. Second, our heuristic layouts outperform the ones produced by OA, especially on Box 1. Looking at the PSR values (shown in parenthesis in Figure 3), we also notice that OA's PSR is only 95% and 90% on Box 1 and Box 2 respectively, while DOT achieves a PSR of 100% in both cases. Third, all the other simple layouts (except the All H-SSD case) have a lower TOC, but lead to longer response times. Finally, looking at the PSR numbers in Figure 3 (shown in the parenthesis) for these simple layouts, we observe that these layout (except All H-SSD) have PSR values that are less than 100% – meaning that some queries in these layouts don't meet the required performance targets.

Figure 4 (a) and (b) shows our heuristic layouts for the Box 1 and Box 2 configurations. In these figures, the primary index associated with a table is denoted by appending the suffix "_pkey" to the table name (e.g. `partsupp` has an primary index file called `partsupp_key`).

From Figure 4, we observe that some table objects (e.g. `lineitem`) that tend to be accessed frequently with the SR I/O requests, are placed on the HDD RAID 0 in **Box 1** and on the L-SSD RAID 0 in **Box 2**. RAID 0 systems are very cost-effective for SR I/O patterns as seen in Table 1: The SSD RAID 0 achieves SR I/O performance comparable to H-SSD (x1.3) with significantly lower storage cost (x0.056). The HDD RAID 0 can be similarly compared with the L-

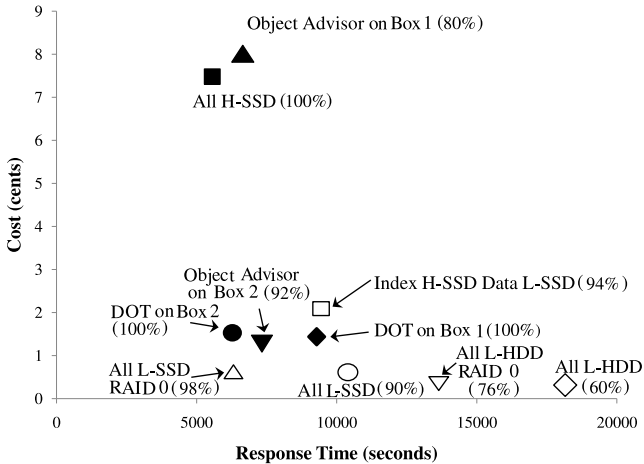

Figure 7: The modified TPC-H workload with relative SLA = 0.25. The number in parenthesis associated with each label indicates the PSR value (%).

SSD (x1.36 faster at only x0.107 of the storage cost). DOT leverages these RAID 0 systems to save on the TOC with only a small (and acceptable) performance penalty.

Notice in Figure 4 that some tables (e.g. `partsupp`) and their primary key indices are still placed on the H-SSD. In fact, some queries (e.g., Query #2) involves RR I/O. Since the performance gap between the H-SSD and the RAID 0 system is large for RR I/O, we still need to put these objects on the H-SSD to meet the (relative) SLA requirements.

We have also repeated the experiment above with the relative SLA value set to 0.25. The heuristic layouts are similar as the ones when the relative SLA is 0.5. and we omit these results in the interest of space.

### 4.4.2 The Modified TPC-H Workload

Figure 5 shows the cost/performance comparison for the different layouts on the modified TPC-H workload, when the relative SLA is 0.5. From the PSR values in this figure (shown in parenthesis in the figure), we observe that all the simple layouts (except the ALL H-SSD case) fail to achieve the target SLA, resulting in low PSR values.

Figure 6 illustrates the layout created by DOT when the relative SLA is 0.5. Observe the difference from the original TPC-H workload experiments (shown in Figure 4): now DOT places most of the data objects on the H-SSD device in both the Box 1 and 2 configurations. For this modified workload we now have more selective predicates, and the query optimizer has more opportunities to exploit the high-performance RR I/O characteristics of the H-SSD device by using indexed NLJ (INLJ). In fact, we have observed that on the DOT layouts (across both box configurations), with this modified TPC-H workload (and relative SLA = 0.5), 50% of the joins in the query plans for this workload are INLJ, whereas only 11% of the joins in the original TPC-H workload (discussed in Section 4.4.1) were INLJ.

Although DOT has to use mostly the H-SSD device to meet the SLA, Figure 5 shows that DOT still saves on the TOC compared to the All H-SSD layout.

Now, we relax the relative SLA to 0.25. The results for this experiment are shown in Figure 7. From this figure, we observe that the TOC with DOT is **5X** lower than the TOC with the All H-SSD layout, while achieving a 100% PSR.

The DOT layouts for in this case are omitted here (but can be found in [3]). We observed that compared to the case when the relative SLA = 0.5 (Figure 6), now some bulk data (e.g. `lineitem`) moves to the cheaper storage classes, such as L-SSD RAID 0 on Box 2, and HDD RAID 0 on Box 1.

Another interesting observation across the two different relative SLAs of 0.5 and 0.25 above, comes from taking a closer look at the ratio of Indexed NLJ (% INLJ) that are used in each case with the DOT layouts on both box configurations. As noted above, with a relative SLA value of 0.5 about 50% of the join operations are INLJs. Looking at the query plans for the case when the relative SLA value is 0.25, we observe that only 33% of the query plans in the DOT configurations (in both box configurations) are now INLJs. As the SLA constraint loosens, DOT moved the data around and switched query plans to use more hash join algorithms (rather than INLJ) to achieve the target SLA. This observation demonstrates the need to consider query optimization along with data layout optimization.

### 4.4.3 Heuristics Versus Exhaustive Search

In this experiment, we compare the heuristics (introduced in Section 3.1) with exhaustive search algorithms in terms of the TOC and performance of the layouts that each method recommends. The Exhaustive Search (ES) method explores all possible layouts and evaluates each one of them using the same TOC and performance estimation as DOT.

To allow the ES method to complete, we use a smaller workload in this experiment. This workload consists of 33 TPC-H queries generated from the 11 TPC-H query templates, which are a subset of the original 22 TPC-H queries template. These queries are: Q1, Q3, Q4, Q6, Q12, Q13, Q14, Q17, Q18, Q19, Q22. The reason why we use a subset of queries is that ES explores an exponential number (i.e., $M^N$) of layouts. If we use the whole TPC-H data set (that contains 16 objects), the number of all possible layout is 43 million, which we estimate would take about 3,500 hours for ES to compute. To make the ES method run in a reasonable amount of time, we use eight TPC-H data objects (`lineitem`, `orders`, `customer`, `part` and their indices) and a subset of TPC-H original queries for this experiment.

In this experiment, we fix the relative SLA to 0.5 and vary the capacity limits on the storage classes to compare the performance of DOT and ES. Adding capacity constraints makes the feasible search space more challenging for the greedy heuristics to explore.

We enforce capacity limits on the HDD RAID 0 and the HDD storage devices. As described in Section 4.4.1, the H-SSD and the L-SSD devices are not heavily used in the original TPC-H queries, so we do not add capacity limits to these devices. We ran a test run of ES on both configurations and found the space that it needs on the HDD Raid 0 device in Box 1 and the HDD device in Box 2, which was 27GB and 8.8GB respectively. Then, we set the capacity limits for these devices to be around these limits, to 24GB and 8GB on Box 1 and Box 2 respectively, and then decrease this limit by half each time.

From this experiment, we observed that DOT's performance (both in terms of the TOC and the response time) is comparable to that of ES – DOT's response time in these experiments was within 9% of ES in all cases, and its TOC was within 16% of ES in most cases. In the interest of space we omit the detailed graphs, and refer the reader to [3].

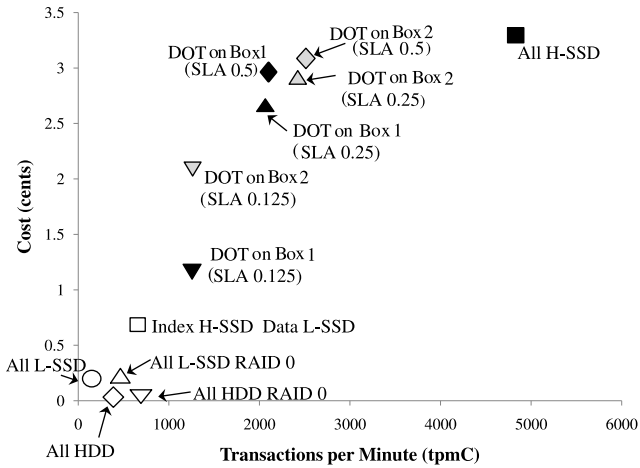

Figure 8: TPC-C results

| | SLA 0.5 | SLA 0.25 | SLA 0.125 |
|---|---|---|---|
| HDD | pk_warehouse pk_customer orders item, pk_item pk_district pk_orders i_orders | pk_warehouse item, pk_item pk_orders pk_district history pk_customer new_order i_orders orders pk_new_order | pk_warehouse item, pk_item pk_customer pk_district new_order pk_new_order orders pk_orders pk_order_line i_orders |
| L-SSD RAID 0 | i_customer | i_customer | customer i_customer |
| H-SSD | customer pk_stock pk_order_line district order_line warehouse stock history new_order pk_new_order | district pk_order_line pk_stock order_line warehouse customer stock | warehouse district pk_stock history order_line stock |

Table 3: DOT layouts under different relative SLAs on Box 2 for the TPC-C workload.

Also, DOT computes the layouts orders of magnitude faster than ES taking about 9 seconds in each case, compared to 1,400 seconds for ES.

### 4.5 TPC-C Experiments

For the TPC-C experiments, we measured and compared the performance of the different layouts on two metrics, the New-Order transactions per minute(**tpmC**) and the TOC. For this experiment, we used the Database Test Suite 2 [1] (DBT2), which is a fair implementation of the TPC-C benchmark, and populated a 30GB (scale factor 300) TPC-C database. DBT2 provides various workload parameters, such as *terminals/warehouse*, *DB connections* and think time. In our experiment, we choose 300 *DB connnections*, 1 *terminals/warehouse*, no think time, set the measurement period of TPC-C workload to 1 hour, and use two minutes to ramp-up the database.

#### 4.5.1 Workload Profiling

We observed that most I/O patterns in the TPC-C workload are *random* accesses, even when all the data objects are placed on the HDD. From this observation, we estimate that the query plans will not change (from random access to sequential access) even if the data objects are moved to HDD. Thus, in this experiment, we only need one simple layout: namely, the All H-SSD case.

To generate the workload profiles (see Section 3.4), we use a *test run* instead of the estimates from the query optimizer, since the TPC-C queries have short latencies, and the test run can give actual I/O statistics. After a 5-minute test run, our layout technique uses the workload profiles and the I/O profiles (estimated under 300 degree of concurrency) to get a TOC-effective layout.

#### 4.5.2 Performance Results

First, we evaluate the effect on the TOC when using DOT with varying performance constraints. We ran the TPC-C workload on both Box 1 and Box 2 with relative SLA values of 0.5, 0.25, 0.125, without capacity limits on any of the storage classes. Here, relative SLA = 0.5 means that the observed tpmC should be higher half of the tpmC that can be achieved with all the data on the H-SSD.

Figure 8 shows the effectiveness of each layout in terms of the tpmC and the TOC. From this figure, we observe that the TOC with DOT decreases as the relative SLA is relaxed: DOT on Box1 with the relative SLA = 0.125 has about **3X** smaller TOC compared to the All H-SSD case.

From the data layout on Box 2 configuration shown in Table 3, we observe that as the relative SLA is relaxed, more objects are shifted from the expensive storage classes to the cheaper ones. (We observed a similar trend for Box 1 – see [3] for details.)

An interesting observation is we saw is that the L-SSD device in Box 1 is seldom used (see [3] for details), since the L-SSD device has poor random write (RW) performance, as seen from Table 1. Even though the L-SSD device is faster than the HDD RAID 0 device for RR I/O, the difference is not big enough to overcome the L-SSD's poor RW I/O and expensive TOC. Therefore, most objects are laid out on the HDD RAID 0 and the H-SSD devices in Box 1. However, for the layout on Box 2 (see Table 3), the `customer` object is placed on the L-SSD RAID 0 device when the relative SLA = 0.125 even though it is accessed frequently using RW I/O (Table 3). The reason for this behavior is that the RAID 0 device can improve random write performance by distributing the write evenly over the two disks. Coupled with RAID 0, the L-SSD device can still be utilized in the TPC-C workloads.

Overall, these results indicate that even with the TPC-C like workload, DOT can produce TOC-efficient data layouts.

#### 4.5.3 Heuristics Versus Exhaustive Search

We also compared the DOT heuristics introduced in Section 3.1 to the Exhaustive Search (ES) for the TPC-C workload. In this experiment, we use the entire TPC-C benchmark, and set the relative SLA = 0.25. We also vary the capacity constraints (as we did for the comparison with ES in the TPC-H case described in Section 4.4.3). In this experiment, we enforce capacity constraints only on the H-SSD, since this device are often the most capacity constrained. The specific capacity constraint values that we use for the H-SSD are: No Limit and 21GB.

Notice that, given the stringent constraints both on the capacity and the performance, there may be no feasible so-lution. In such a case, we slightly relax the relative SLA and repeat the optimization as illustrated in Figure 2. This process stops when ES finds a feasible solution.

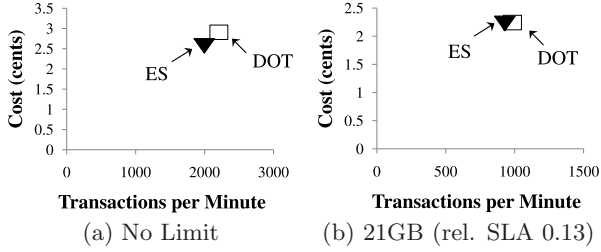

**Figure 9: ES vs DOT with different capacity limits on Box 2 for the TPC-C workload.**

The result for this experiment for Box 2 is shown in Figure 9. Each graph is associated with the capacity limit on the H-SSD device, and the final relative SLA value. ES and DOT achieve almost same result (tpmC and TOC). In this case, DOT computes the layouts in 3 seconds compared to 800 seconds for ES.

We have also run the experiment above with relative SLA values of 0.5 and 0.125, on both Box 1 and Box 2, and capacity limits of 18GB, 15GB and 12GB. The results presented above are representative of the results in these other cases (with DOT and ES having nearly the same TOC and tpmC performance).

## 5. DISCUSSIONS

In this section, we discuss two possible extensions of this work: namely, (1) other possible problem definitions, and (2) the discrete-sized storage cost model. In the extended version of this paper [3], we run extensive experiments to verify our DOT method can work well with these extensions.

### 5.1 Other Problem Formulations

An interesting complementary problem to the one that we use in this paper is to pick the "right" server hardware from a range of options, for a pre-defined workload. In the following, we formally define this problem. In [3], we experimentally show that given some options about the storage configurations, our DOT method is able to recommend the TOC-efficient storage configuration and the data layout, while guaranteeing the SLAs of the input workloads.

#### 5.1.1 The Generalized Provisioning Problem

**Input**: (1) Database objects $O = \{o_1, \cdots, o_N\}$, (2) Storage Configurations Options $F = \{f_1, \cdots, f_X\}$, where each $f_i$ uses the storage classes $D_i = \{d_1^i, \cdots, d_M^i\}$ with price (TOC/GB/hour) $P_i = \{p_1^i, \cdots, p_M^i\}$ and capacity $C_i = \{c_1^i, \cdots, c_M^i\}$, (3) Query workload $W = \{[q_1^1, \cdots, q_n^1], \cdots, [q_1^c, \cdots, q_n^c]\}$ with performance constraints $T = \{t_i^j\}$.

**Output**: A storage configuration $f_k$ with the data layout $L_k$ on $f_k$: $O \to D_k$ that minimizes the TOC $C(L_k, W) = C(L_k) * t(L_k, W)$ for a given $W$ where

$$C(L_k) = \Sigma_{d_j^k \in D_k}(p_j^k * S_j^k)$$

under the capacity constraints, $\Sigma_{o_i \in O_j^k} s_i < c_j^k$ ($j = 1, \cdots, M$), and performance constraints $T = \{t_i^j\}$.

The extended version of this paper [3], contains experimental results that show how the DOT framework can be extended to solve this generalized provisioning problem.

### 5.2 Discrete-sized Storage Cost Model

In Section 2.1, we define the **layout cost** as $C(L) = \Sigma_{d_j \in D}(p_j * S_j)$, where $C(L)$ is in linear relationship with the actual space usage $S_j$ on $d_j$. However, the storage devices are generally bought in discrete-sized units (e.g. 40GB, 80GB, 120GB) so $C(L)$ may not vary linearly with $S_j$. To adapt to this discrete-sized case, we generalize our **layout cost** definition as follows.

**Layout cost** *(cent/hour)*: Assume that a database is laid out on $D$, taking $S_j$ GB space for each storage class $d_j$ ($S_j \geq 0$). The price and capacity of $d_j$ are $p_j$ and $c_j$ respectively. Now, let $L$ denote this particular layout. Then, the cost per hour for this layout $L$, denoted as $C(L)$, is computed as:

$$C(L) = \Sigma_{d_j \in D}[\alpha * (p_j * c_j) + (1 - \alpha) * (S_j/c_j) * (p_j * c_j)]$$

As seen from the above formula, the layout cost $C(L)$ is composed of two parts: $(p_j * c_j)$ and $(S_j/c_j) * (p_j * c_j)$. The first part, namely $(p_j * c_j)$, is the **discrete cost** determined by the number of identical devices in a certain storage class. The discrete cost has to be paid no matter how much space is used in that storage class. On the other hand, $(S_j/c_j) * (p_j * c_j)$ is the **linear cost** determined by the proportional space usage. The variable $\alpha$ is a tunable parameter that can adjust the weights between the discrete and the linear costs.

In the extended version of this paper [3], we include more discussions on this cost model and present experimental results demonstrating that our DOT method still works with this storage cost model.

## 6. RELATED WORK

The problem of data placement involves assigning $N$ data objects to $M$ storage devices with the objective of improving the workload performance. A recent work on this topic by Koltsidas *et al.* [18] examines the optimal data page placement between a SSD and a traditional HDD. They propose a family of online buffer pool replacement algorithm so that pages are placed on the right devices for better workload performance. Canim *et al.* [10] propose an Object Advisor to place database objects (e.g. tables or indices) on either SSDs or HDDs. Their method first collects the I/O statistics of a workload and then uses a greedy algorithm to decide the placement of the tables and indices. Our work differs from this work in many aspects. First, their goal is to maximize the workload performance by using a limited capacity on the SSDs, while our goal is to minimize the TOC that is incurred when running that workload. Second, their query optimizer is not aware of the specific characteristics of the SSDs, so they miss the interactions between the query plans and data layouts. In contrast, we design and employ an extended query optimizer in Section 3.5 to make the query optimizer aware of different storage classes, and our method is able to update the cheapest query plan dynamically as the data layout is changed.

The virtual machine placement problem, as proposed in [7, 12, 17], is to deploy the virtual machines on the most suitable physical hosts for either better resource utilization of the hosts, lower operating costs, or load balancing among hosts. The connection between the virtual machine placement problem and our work is that the problem they want to solve is to find a mapping of a virtual machine onto physical hosts, while our problem is to find a mapping (or data layout) between data objects and storage classes. The commonality at a high level is that both consider performance constraints (e.g. SLA) and that the naive solution space size in both cases is exponential and impractical to explore all possible solutions. But besides that high-level similarity, their problem and our problem are very different in terms of the goal and solution.

Another branch of related work is *index advisor*, or alternatively, *physical design tuning* [6,8,9,13]. The only similarity between the *index advisor* problem and our problem is we both consider the impact of indexes on query execution within the storage bounds. However, the differences between these two problems are significant. First, their problem statement is: given all possible index selection choices, determine which indexes should be selected to create. While our problem assumes that the indexes are already selected and we consider *which storage class each index and data should be placed on*. Other important difference is that their objective is to maximize the query/workload performance, while our goal is to minimize the total operation cost, while maintaining a certain level of the query performance.

With the maturity of SSDs, substantial research has focused on improving the DBMS performance by using SSDs including revisiting the five-minute rule based [15], examining methods for improving various DBMS internals such as query processing techniques [25, 27], index structure [5, 21, 24], bufferpool extension [11], page layout [19] and temporary space [20]. These methods are complementary to our work here, as these efforts allow the DBMS to use SSDs more effectively – hence, these methods can be easily used along with our method in a DBMS that is tuned for SSDs.

# 7. CONCLUSIONS AND FUTURE WORK

This paper has introduced a new problem of provisioning I/O resources for a workload to minimize the total operating cost that is incurred when running that workload. This paper has also presented the design of a solution, called DOT, for this problem. DOT extends the query optimization components that are already present in a modern DBMS, and hence is a practical solution. We have implemented DOT in PostgreSQL, and have presented extensive evaluation of DOT using various DSS and OLTP workloads. These results demonstrate that DOT can produce significant reductions in total operation costs, while meeting performance SLAs.

There is a wide range of future work that is possible, including examining every aspect of database query processing, database query optimization, physical database design, etc., from the new perspective of minimizing the total operating cost while meeting traditional performance targets set in SLAs. Other future work includes extending the DOT framework to help make purchasing and capacity planning decisions; for example, by running DOT iteratively to determine the TOC and SLA performance of different hardware configurations under consideration.

# 8. ACKNOWLEDGMENTS

The work by the first author was primarily done while the author was at the NEC Laboratories of America. This research was supported in part by a grant from the National Science Foundation under grant IIS-0963993, and a gift donation from the NEC Laboratories of America.